\documentclass[11pt]{llncs}
\usepackage{epsfig}
\usepackage{wrapfig}
\usepackage{fancyvrb}
\usepackage{subfigure}
\usepackage{latexsym}
\usepackage{pslatex}
\usepackage{amssymb}
\begin{document}

%%%%%%%%%%%%%%%%%%%%%%%%%%%%%%%%%%%%%%%%%%%%%%%%%%%%%%%%%%%%%%%%%%%%%%%%%%%%%%%%

\newcommand{\vipress}{{\tt ViPReSS} }
\newcommand{\vipressnospace}{{\tt ViPReSS}}
\newcommand{\btw}{{\sf BTW} }
\newcommand{\btwnospace}{{\sf BTW}}
\newcommand{\codeframesep}{1mm}
\newcommand{\eat}[1]{}

%%%%%%%%%%%%%%%%%%%%%%%%%%%%%%%%%%%%%%%%%%%%%%%%%%%%%%%%%%%%%%%%%%%%%%%%%%%%%%%%

%     * Page Layout
%           o \columnsep: gap between columns
%           o \topmargin: gap above header
%           o \topskip: between header and text
%           o \textheight: height of main text
%           o \textwidth: width of text
%           o \oddsidemargin: odd page left margin
%           o \evensidemargin : even page left margin
%     * Paragraphs
%           o \parindent: indentation of paragraphs
%           o \parskip: gap between paragraphs
%     * Floats (tables and figures)
%           o \floatsep: space left between floats.
%           o \textfloatsep: space between last top float or first bottom float and the text.
\addtolength{\textfloatsep}{-5mm}
%           o \intextsep : space left on top and bottom of an in-text float.
\addtolength{\intextsep}{-5mm}
%           o \dbltextfloatsep is \textfloatsep for 2 column output.
%           o \dblfloatsep is \floatsep for 2 column output.
%           o \abovecaptionskip: space above caption
\addtolength{\abovecaptionskip}{-8mm}
%           o \belowcaptionskip: space below caption
\addtolength{\belowcaptionskip}{-3mm}
%     * Maths
%           o \abovedisplayskip: space before maths
\addtolength{\abovedisplayskip}{-5mm}
%           o \belowdisplayskip: space after maths
\addtolength{\belowdisplayskip}{-5mm}
%           o \arraycolsep: gap between columns of an array
%     * Lists
%           o \topsep: space between first item and preceding paragraph.
\setlength{\topsep}{0mm}
%           o \partopsep: extra space added to \topsep when environment starts a new paragraph.
\setlength{\partopsep}{0mm}
%           o \itemsep: space between successive items.
% 
% 
%%	p.6 eerste zin - het klinky een beetje raar...
%%	    ``stratified'' -> is het duidelijk of moet een woordje uitleg bij?
%%	p.8-9 in/out. Iets zeggen zoals ``This refactoring can be/should be/...
%%	supported by mode analysis'' + referentie(s)? Ik krijg een beetje het
%%	gevoel dat de mensen van programmatransformatie onze inzending gaan lezen
%%	en wij moeten de juiste referenties op de juiste plaatsen hebben om te
%%	tonen dat wij ons domein goed kennen.
%%	
%%	Ik zou hier ook iets toevoegen over het vervangen van \+ X = Y (or X\=Y)
%%	door meer specifieke vergelijkingen zoals \== of =\=.

%%%%%%%%%%%%%%%%%%%%%%%%%%%%%%%%%%%%%%%%%%%%%%%%%%%%%%%%%%%%%%%%%%%%%%%%%%%%%%

\title{Improving Prolog Programs: Refactoring for Prolog}

\author{Tom Schrijvers\thanks{Research Assistant of the Fund for Scientific Research-Flanders (Belgium)(F.W.O.-Vlaanderen)}\inst{1}
\and Alexander Serebrenik\thanks{The research presented has been carried out during the second author's stay at Department of Computer Science, K.U. Leuven, 
Belgium and STIX, {\'E}cole Polytechnique, France}\inst{2}}
\institute{Department of Computer Science, K.U. Leuven\\
Celestijnenlaan 200A, B-3001, Heverlee,
Belgium\\E-mail: Tom.Schrijvers@cs.kuleuven.ac.be
\and
Laboratory of Quality of Software (LaQuSo), T.U. Eindhoven\\
HG 5.91, Den Dolech 2, P.O. Box 513, 5600 MB  Eindhoven,
The Netherlands\\E-Mail: a.serebrenik@laquso.com}

\maketitle

%%%%%%%%%%%%%%%%%%%%%%%%%%%%%%%%%%%%%%%%%%%%%%%%%%%%%%%%%%%%%%%%%%%%%%%%%%%%%%%%
\begin{abstract}

{\em Refactoring} is an established technique from the OO-community to
restructure code: it aims at improving software readability, maintainability
and extensibility. Although refactoring is not tied to the OO-paradigm in
particular, its ideas have not been applied to Logic Programming until now.

This paper applies the ideas of refactoring to Prolog programs.  A catalogue
is presented listing refactorings classified according to scope. Some of
the refactorings have been adapted from the OO-paradigm, while others have been
specifically designed for Prolog. Also the discrepancy between intended
and operational semantics in Prolog is addressed by some of the refactorings.

In addition, \vipressnospace, a semi-automatic refactoring browser,
is discussed and the experience with applying \vipress to a large Prolog
legacy system is reported. Our main conclusion is that refactoring is not only
a viable technique in Prolog but also a rather desirable one.

\end{abstract}

%%%%%%%%%%%%%%%%%%%%%%%%%%%%%%%%%%%%%%%%%%%%%%%%%%%%%%%%%%%%%%%%%%%%%%%%%%%%%%%%
%%%\section{Overview of Refactoring}
\section{Introduction}\label{sec:refactoring}

Program changes take up a substantial part of the entire programming
effort. Often changes are required to incorporate additional functionality
or to improve efficiency. In both cases, a preliminary step of improving
the design without altering the external behaviour is recommended. This
methodology, called {\em refactoring}, emerged from a number of pioneer results
in the OO-community~\cite{Fowler:et:al,Opdyke:PhD,Roberts:Brant:Johnson} and
recently came to prominence for functional languages~\cite{Li:Reinke:Thompson}.
More formally, refactoring is a source-to-source program transformation that
changes program structure and organisation, but not program functionality.
The major aim of refactoring is to improve readability, maintainability
and extensibility of the existing software. While performance improvement
is not considered as a crucial issue for refactoring, it can be noted
that well-structured software is more amenable to performance tuning. We
also observe that certain techniques that were developed in the context of
program optimisation, such as dead-code elimination and redundant argument 
filtering, can improve program organisation and, hence, can be
considered refactoring techniques. In this paper we discuss additional
refactoring techniques for Prolog programs.

To achieve the above goals two questions need to be answered:  {\em where}
and {\em how} transformations need to be performed. Unlike automated program
transformations, neither of the steps aims at transforming the program fully
automatically. The decision whether to transform is left to the program
developer. However, providing automated support for refactoring is useful
and an important challenge.

Deciding automatically {\em where} to apply a transformation can be a difficult
task on its own. Several ways to resolve this may be considered.  First,
program analysis approaches can be used. For example, it is common practice
while ordering predicate arguments to start with the input arguments and end
with the output arguments. Mode information can be used to detect when this
rule is violated and to suggest the user to reorder the arguments. Second,
machine learning techniques can be used to predict further refactorings
based on those already applied. Useful sequences of refactoring steps can be
learned analogously to automated macro construction~\cite{Jacobs:Blockeel}.
Following these approaches, automatic refactoring tools, so called {\em
refactoring browsers}, can be expected to make suggestions on where refactoring
transformations should be applied. These suggestions can then be either
confirmed or rejected by the program developer.

Answering {\em how} the program should be transformed might also require the
user's input. Consider for example a refactoring that renames a predicate:
while automatic tools can hardly be expected to guess the new predicate name,
they should be able to detect all program points affected by the change.
Other refactorings require certain properties, like as absence of user-defined
meta-predicates, that cannot be easily inferred. It is then up to the user
to evaluate whether the properties hold.

%%%%%%%%%%%%%%%%%%%%%%%%%%%%%%%%%%%%%%%%%%%%%%%%%%%%%%%%%%%%%%%%%%%%%%%%%%%%%%%%
%% EX-Introduction

The outline of this paper is as follows. We first
illustrate the use of several refactoring techniques on a small example in
Section \ref{sec:example}. Then a more comprehensive catalogue of Prolog refactorings
is given in Section \ref{sec:catalogue}. In Section \ref{sec:vipress}
we introduce \vipress, our refactoring browser, currently implementing most of
the refactorings of the catalogue. \vipress has been successfully applied for
refactoring a 50,000 lines-long legacy system. Finally,
in Section \ref{sec:conclusions} we conclude.
\vspace{-9mm}
%%%%%%%%%%%%%%%%%%%%%%%%%%%%%%%%%%%%%%%%%%%%%%%%%%%%%%%%%%%%%%%%%%%%%%%%%%%%%%%%
\section{Detailed Prolog Refactoring Example}\label{sec:example}

We illustrate some of the techniques proposed by a detailed refactoring example.
Consider the following code fragment borrowed from O'Keefe's ``The Craft of
Prolog'' \cite{OKeefe}, p. 195. It describes three operations on a {\em reader}
data structure used to sequentially read terms from a file. The three
operations are \texttt{make\_reader/3} to initialise the data structure,
\texttt{reader\_done/1} to check whether no more terms can be read and
\texttt{reader\_next/3} to get the next term and advance the reader.

\begin{Verbatim}[commandchars=\\\{\},frame=single,fontsize=\small,framesep=\codeframesep,label=O'Keefe's original version]
\textbf{make_reader}(File,Stream,State) :-
        open(File,read,Stream),
        read(Stream,Term),
        reader_code(Term,Stream,State).

\textbf{reader_code}(end_of_file,_,end_of_file) :- ! .
\textbf{reader_code}(Term,Stream,read(Term,Stream,Position)) :-
        stream_position(Stream,Position).

\textbf{reader_done}(end_of_file).

\textbf{reader_next}(Term,read(Term,Stream,Pos),State)) :-
        stream_position(Stream,_,Pos),
        read(Stream,Next),
        reader_code(Next,Stream,State).
\end{Verbatim}
\vspace{-3mm}

We will now apply several refactorings to the above program to improve
its readability.

First of all, we use if-then-else introduction to get rid of the ugly red
cut in the \texttt{reader\_code/3} predicate:

% if then else

\begin{Verbatim}[commandchars=\\\{\},frame=single,fontsize=\small,framesep=\codeframesep,label=Replace cut by if-then-else]
\textbf{reader_code}(Term,Stream,State) :-
        \underline{(} Term = end_of_file,
          State = end_of_file \underline{->}
                true
        \underline{;}
                State = read(Term,Stream,Position),
                stream_position(Stream,Position)
        \underline{)}.
\end{Verbatim}
\vspace{-3mm}

This automatic transformation reveals two malpractices, the first of which is
producing output before the commit, something O'Keefe himself disapproves of
(p. 97). This is fixed manually to:

% manual clean-up 1

\begin{Verbatim}[commandchars=\\\{\},frame=single,fontsize=\small,framesep=\codeframesep,label=Output after commit]
\textbf{reader_code}(Term,Stream,State) :-
        ( Term = end_of_file ->
                \underline{State = end_of_file}
        ;
                State = read(Term,Stream,Position),
                stream_position(Stream,Position)
        ).
\end{Verbatim}
\vspace{-3mm}

The second malpractice is a unification in the condition
of the if-then-else where actually an equality test is meant.
Consider that the \texttt{Term} argument is a variable. Then
the binding is certainly unwanted behaviour. Manual change
generates the following code:

% manual clean-up 2

\begin{Verbatim}[commandchars=\\\{\},frame=single,fontsize=\small,framesep=\codeframesep,label=Equality test]
\textbf{reader_code}(Term,Stream,State) :-
        ( \underline{Term == end_of_file} ->
                State = end_of_file
        ;
                State = read(Term,Stream,Position),
                stream_position(Stream,Position)
        ).
\end{Verbatim}
\vspace{-3mm}

Next, we notice that the sequence \texttt{read/2, reader\_code/3}
occurs twice, either by simple observation or by computing common body
subsequences. By applying predicate extraction of this common
sequence, we get:

% predicate extraction

\begin{Verbatim}[commandchars=\\\{\},frame=single,fontsize=\small,framesep=\codeframesep,label=Predicate extraction]
\textbf{make_reader}(File,Stream,State) :-
        open(File,read,Stream),
        \underline{read_next_state(Stream,State)}.

\textbf{reader_next}(Term,read(Term,Stream,Pos),State)) :-
        stream_position(Stream,_,Pos),
        \underline{read_next_state(Stream,State)}.

\underline{\textbf{read_next_state}(Stream,State) :-}
        \underline{read(Stream,Term),}
        \underline{reader_code(Term,Stream,State).}
\end{Verbatim}
\vspace{-3mm}

Next we apply O'Keefe's own principle of putting the input argument first
and the output arguments last (p. 14--15):

% reorder arguments

\begin{Verbatim}[commandchars=\\\{\},frame=single,fontsize=\small,framesep=\codeframesep,label=Argument reordering]
\textbf{reader_next}\underline{(read(Term,Stream,Pos),Term,State)} :-
        stream_position(Stream,_,Pos),
        read_next_code(Stream,State).
\end{Verbatim}
\vspace{-3mm}

Finally, we introduce less confusing and overlapping names for the
\texttt{read/3} functor, the \texttt{stream\_position/[2,3]} built-ins and
a more consistent naming for \texttt{make\_reader}, more in line with the other
two predicates in the interface. O'Keefe stresses the importance of
consistent naming conventions (p. 213).

Note that direct renaming of built-ins such as \texttt{stream\_position}
is not possible, but a similar effect can be achieved by
extracting the built-in into a new predicate with the desired name.

% renaming

%\vspace{1cm}

\begin{Verbatim}[commandchars=\\\{\},frame=single,fontsize=\small,framesep=\codeframesep,label=Renaming]
\underline{\textbf{reader_init}}(File,Stream,State) :-
        open(File,read,Stream),
        reader_next_state(Stream,State).

\textbf{reader_next}(\underline{reader}(Term,Stream,Pos),Term,State)) :-
        \underline{set_stream_position}(Stream,Pos),
        reader_next_state(Stream,State).

\textbf{reader_done}(end_of_file).

\textbf{reader_next_state}(Stream,State) :-
        read(Stream,Term),
        build_reader_state(Term,Stream,State).

\textbf{build_reader_state}(Term,Stream,State) :-
        ( Term == end_of_file ->
                State = end_of_file
        ;
                State = \underline{reader}(Term,Stream,Position),
                \underline{get_stream_position}(Stream,Position)
        ).

\underline{\textbf{set_stream_position}(Stream,Position) :-}
        \underline{ stream_position(Stream,_,Position).}
\underline{\textbf{get_stream_position}(Stream,Position) :-}
        \underline{ stream_position(Stream,Position).}
\end{Verbatim}
\vspace{-3mm}

While the above changes can be performed manually, a refactoring browser such as
\vipress (see Section \ref{sec:vipress}) guarantees consistency, correctness
and furthermore can automatically single out opportunities for refactoring.
\vspace{-3mm}
%%%%%%%%%%%%%%%%%%%%%%%%%%%%%%%%%%%%%%%%%%%%%%%%%%%%%%%%%%%%%%%%%%%%%%%%%%%%%%%%
\section{Comprehensive Catalogue of Prolog refactorings}\label{sec:catalogue}

\vspace{-3mm}
In this section we present a number of refactorings that we have found to be
useful when Prolog programs are considered. A more comprehensive discussion
of the presented refactorings can be found in~\cite{Schrijvers:Serebrenik:Demoen}.

We stress that the programs are
not limited to pure logic programs, but may contain various built-ins such as
those defined in the ISO standard \cite{ISO13211-1}. The only exception are 
higher-order constructs that are not dealt with automatically, but manually.
Automating the detection and handling of higher-order predicates is an
important part of future work.

The refactorings in this catalogue are grouped by scope. The scope expresses
the user-selected target of a particular refactoring. While the particular
refactoring may affect code outside the selected scope, it is only because
the refactoring operation detects a dependency outside the scope. 

For Prolog programs we distinguish the following four scopes, based on the
code units of Prolog: system scope (Section \ref{sub:system}), module scope
(Section \ref{sub:module}), predicate scope (Section \ref{sub:predicate})
and clause scope (Section \ref{sub:clause}).
\vspace{-4mm}
%%%%%%%%%%%%%%%%%%%%%%%%%%%%%%%%%%%%%%%%%%%%%%%%%%%%%%%%%%%%%%%%%%%%%%%%%%%%%%%%
\subsection{System Scope Refactorings}\label{sub:system}

The system scope encompasses the entire code base. Hence the user
does not want to transform a particular subpart, but to affect the
system as a whole.

% %\vspace{1mm}
% \renewcommand{\FancyVerbFormatLine}[1]{%
%         \makebox[3mm][l]{\textbullet}#1}
% \begin{Verbatim}[commandchars=\\\{\},frame=single,fontfamily=\familydefault,framesep=\codeframesep,label=System Scope Refactorings,xleftmargin=1cm,xrightmargin=1cm]
% Extract common code into predicates
% Hide predicates (remove them from export lists)
% Remove dead code
% Remove duplicate predicates
% Remove redundant arguments
% Rename functor
% \end{Verbatim}
\vspace{-4mm}
\subsubsection{Extract common code into predicates}

This refactoring looks for common functionality across the system
and extracts it into new predicates. The common functionality consists
of subsequences of goals that are called in different predicate bodies. By
replacing these common subsequences with calls to new predicates the overall
readability of the program improves. % The affected predicate bodies get
% shorter and the calls to the new predicates can be more meaningful than what
% they replace. 
Moreover the increased sharing simplifies maintenance as now only one copy
needs to be modified.  User input is required to decide what common sequences
form meaningful new predicates. Finding the common sequences and the actual
replacing are handled automatically by \vipressnospace.
\vspace{-5mm}
\subsubsection{Hide predicates}

This refactoring removes export declarations for predicates that are not
imported in any other module. User input is required to confirm that a
particular predicate is not meant for use outside the module in the future.
This refactoring simplifies the program by reducing the number of entry
points into modules and hence the intermodule dependencies.
\vspace{-5mm}
\subsubsection{Remove dead code}

Dead code elimination is sometimes performed in compilers for efficiency
reasons, but it is also useful for developers: dead code clutters the program.

We consider a predicate definition in its entirety as a code unit that can
be dead, as opposed to a subset of clauses. While eliminating a subset of 
clauses can change the semantics of the predicate and hence lead to an 
erroneous use, this is not the case if the entire predicate is removed. 

It is well-known that reachability of a certain program point (predicate) is,
in general, undecidable. However, one can safely approximate the dead code
by inspecting the {\em predicate dependency graph} (PDG) of the system. The
PDG connects definitions of predicates to the predicates that use them in
their own definition. This graph is useful for other refactorings, like {\em
remove redundant arguments}.  In the system one or more predicates should be
declared as top-level predicates that are called in top-level queries and
form the main entry points of the system.  Now dead predicates are those
predicates not reachable from any of the top-level predicates in the PDG.

User input is necessary whether a predicate can safely be removed or should
stay because of some intended future use.

In addition to unused predicate definitions, redundant predicate import
declarations should also be removed. 
%%%Multiple updates to a module may leave
%%%obsolete import declarations. A module should only import
%%%those predicates from other modules it genuinely needs. 
This may enable the
{\em hide predicate} refactoring to hide more predicates. Dead-code elimination
is supported by \vipressnospace.
 
\vspace{-5mm}
\subsubsection{Remove duplicate predicates}

Predicate duplication or cloning is a well-known problem. One of the
prominent causes is the practice known as ``copy and paste''. Another
cause is unawareness of available libraries and exported predicates in
other modules. The main problem with this duplicate code is its bad
maintainability. Changes to the code need to be applied to all copies.

Looking for all possible duplications can be quite expensive.
% : given $n$
% predicates there are $B_n$ ways\footnote{$B_n$ is the $n^{th}$ Bell number, or the
% number of partitions of a set of size $n$: 
%                 $B_n = \left\lceil e^{-1} \sum_{m=1}^{2n}{m^n \over m!}  \right\rceil $
% .} 
%which predicates can be duplicates of each other. 
In practice in \vipress we limit the number of possibilities by only
considering predicates with identical names in different modules as possible
duplicates. The search proceeds stratum per stratum upwards in the stratified
PDG. In each stratum the strongly connected components
(SCCs) are compared with each other. If all the predicate definitions in an
SCC are identical to those in the other component and they depend on duplicate
components in lower strata, then they are considered duplicates as well.

It is up to the user to decide whether to throw away some of the duplicates
or replace all the duplicate predicates by a shared version in a new module.

\vspace{-5mm}
\subsubsection{Remove redundant arguments}

The basic intuition here is that parameters that are no longer used
by a predicate should be dropped. This problem has been studied,
among others, by Leuschel and S{\o}rensen \cite{Leuschel:Sorensen} in
the context of program specialisation. They
established that the
redundancy property is undecidable and suggested two techniques to find
safe and effective approximations: top-down goal-oriented RAF and bottom-up
goal-independent FAR. In the context of refactoring FAR is the more useful
technique. Firstly, FAR is the only possibility if exported predicates are
considered. Secondly, refactoring-based software development regards the
development process as a sequence of small ``change - refactor - test''
steps. These changes most probably will be local. Hence, FAR is
the technique applied in \vipressnospace.

%%%FAR marks an argument position in the head of the clause as unused if it is
%%%occupied by a variable that appears exactly once in the argument position
%%%that has not been marked as unused. The marking process proceeds bottom-up
%%%per strongly connected component (SCC) of the predicate dependency graph.

The argument-removing technique should consist of two steps. First, unused
argument positions are marked by FAR. 
Second, depending on user input, marked argument
positions are dropped. Similarly to removing unused predicates (dead code
elimination) by removing unused argument positions from predicates we improve
readability of the existing code. 

\vspace{-5mm}
\subsubsection{Rename functor}

This refactoring renames a term functor across the system. If the functor has
several 
different meanings and only one should be renamed, it is up to the user to
identify what use corresponds with what meaning. In a typed language, a meaning
would correspond with a type and the distinction could be made automatically.
Alternatively, type information can be inferred and the renaming can be based
on it.

%%%%%%%%%%%%%%%%%%%%%%%%%%%%%%%%%%%%%%%%%%%%%%%%%%%%%%%%%%%%%%%%%%%%%%%%%%%%%%%%
\subsection{Module Scope Refactorings}\label{sub:module}

The module scope considers a particular module. Usually a module is
implementing a well-defined functionality and is typically contained in
one file.  % This scope corresponds to refactorings operating on classes in
% Java or modules in Haskell.

% \begin{Verbatim}[commandchars=\\\{\},frame=single,fontfamily=\familydefault,framesep=\codeframesep,label=Module Scope Refactorings,xleftmargin=1cm,xrightmargin=1cm]
% Merge modules
% Remove dead code intra-module
% Rename module
% Split module
% \end{Verbatim}
% \renewcommand{\FancyVerbFormatLine}[1]{#1}

\vspace{-5mm}
\subsubsection{Merge Modules}

Merging a number of modules in one can be advantageous in case of strong
interdependency of the modules involved. Refactoring browsers are expected to
discover interrelated modules by taking software metrics such as the number of
mutually imported predicates into account. 
%%%, to suggest the modules to be merged.
Upon user confirmation the actual transformation can be performed.

The inverse refactoring, {\em Split Modules}, is useful to split unrelated
modules or make a large module more manageable.

\vspace{-5mm}
\subsubsection{Remove dead code intra-module}

Similar to {\em dead code removal} for an entire system  (see Section
\ref{sub:system}), this refactoring works at the level of a single module.
It is useful for incomplete systems or library modules with an unknown number
of uses. The set of top level predicates is extended with, or replaced by,
the exported predicates of the module.
\vspace{-5mm}

\subsubsection{Rename module}

This refactoring applies when the name of the module no longer corresponds
to the functionality it implements. %%%%, e.g. due to other refactorings. 
It also
involves updating import statements in the modules that depend on the module.
\vspace{-5mm}

% \subsubsection{Split module}
% 
% This refactoring is the opposite of \texttt{Merge Modules}. By splitting
% a large module into separate modules, the code units become more
% manageable. Moreover, it is easier to reuse a particular functionality if
% it is contained in a separate module. Similarly to the previous refactoring, 
% this one involves updating dependent import statements.

%%%%%%%%%%%%%%%%%%%%%%%%%%%%%%%%%%%%%%%%%%%%%%%%%%%%%%%%%%%%%%%%%%%%%%%%%%%%%%%%
\subsection{Predicate Scope Refactorings}\label{sub:predicate}

The predicate scope targets a single predicate. The code that
depends on the predicate may need updating as well. But this is considered
an implication of the refactoring of which either the user is alerted or
the necessary transformations are performed implicitly.

% \renewcommand{\FancyVerbFormatLine}[1]{%
%         \makebox[3mm][l]{\textbullet}#1}
% \begin{Verbatim}[commandchars=\\\{\},frame=single,fontfamily=\familydefault,framesep=\codeframesep,label=Predicate Scope Refactorings,xleftmargin=1cm,xrightmargin=1cm]
% Add argument
% Move predicate
% Rename predicate
% Reorder arguments
% \end{Verbatim}
% \renewcommand{\FancyVerbFormatLine}[1]{#1}

\vspace{-5mm}
\subsubsection{Add argument}

This refactoring should be applied when a callee needs more information 
from its (direct or indirect) caller. Our experience suggests that the 
situation is very common while developing Prolog programs. It can be 
illustrated by the following example:
%%%\begin{example}
%%%Consider the following program.
\begin{Verbatim}[commandchars=\\\{\},frame=single,framesep=\codeframesep,label=Original Code,fontsize=\small]
compiler(Program,CompiledCode) :-
        translate(Program,Translated),
        optimise(Translated,CompiledCode).

optimise([assignment(Var,Expr)|Statements],CompiledCode) :-
        optimise_assignment(Expr,OptimisedExpr), ...
...
optimise([if(Test,Then,Else)|Statements],CompiledCode) :-
        optimise_test(Test,OptimisedTest), ...

optimise_test(Test,OptimisedTest) :- ...
\end{Verbatim}
Assume that a new analysis ({\tt analyse}) of if-conditions has been 
implemented. Since this analysis requires the original program code
as an input, the only place to plug the call to {\tt analyse} is in
the body of {\tt compiler}:
\begin{Verbatim}[commandchars=\\\{\},frame=single,framesep=\codeframesep,label=Extended Code,fontsize=\small]
compiler(Program,CompiledCode) :-
        analyse(Program,AnalysisResults),
        translate(Program,Translated),
        optimise(Translated,CompiledCode).
\end{Verbatim}
In order to profit from the results of {\tt analyse} the variable
{\tt AnalysisResults} should be passed all the way down to 
{\tt optimise\_test}. In other words, an extra argument should be 
added to {\tt optimise} and {\tt optimise\_test} and its value should be
initialised to {\tt AnalysisResults}.
%%%% $\hfill\Box$\end{example}

Hence, given a variable in the body of the caller and the name 
of the callee, the refactoring browser should propagate
this variable along all possible computation paths from the caller to the
callee. This refactoring is an important preliminary step preceding 
additional functionality integration or efficiency improvement.
\vspace{-5mm}
\subsubsection{Move predicate}
This refactoring 
corresponds to the ``move method'' refactoring of
Fowler~\cite{Fowler:catalogue}. Moving predicate from one module to another
can improve the overall structure of the program by bringing together
interdependent or related predicates.
\vspace{-5mm}
\subsubsection{Rename predicate}
This is the counterpart of the ``rename method'' refactoring.  It can improve
readability and should be applied when the name of a predicate does not reveal
its purpose. Renaming a predicate requires updating the calls to it as well as
the interface between the defining and importing modules.
\vspace{-5mm}
\subsubsection{Reorder arguments}
Our experience suggests that while writing predicate definitions Prolog
programmers tend to begin with the input arguments and to end with the output
arguments. This methodology has been identified as a good practice and even
further refined by O'Keefe \cite{OKeefe} to more elaborate rules.
%%%
%%% Unfortunately, this practice can be violated when additional arguments
%%% are added later. We observed that failure to confirm to this ``input first
%%% output last'' expectation pattern is experienced as very confusing. 
Hence, to improve readability, argument reordering is recommended:
given the predicate name and the intended order of the arguments,
the refactoring browser should produce the code such that the arguments of
the predicate have been appropriately reordered.

It should be noted that most Prolog systems use indexing on the first
argument. Argument reordering can improve the efficiency of the program
execution in this way.

Another efficiency improvement is possible. Consider the fact \texttt{f(a\_out,b\_in)}.
For the query \texttt{?- f(X,c\_in)}, first the variable \texttt{X} is
bound to \texttt{a\_out} and then the unification of \texttt{c\_in} with
\texttt{b\_in} fails. It is more efficient to first unify the input argument
and only if that succeeds bind the output argument. This is somewhat similar
to {\em produce output before commit} in the next section.
\vspace{-5mm}

%%%%%%%%%%%%%%%%%%%%%%%%%%%%%%%%%%%%%%%%%%%%%%%%%%%%%%%%%%%%%%%%%%%%%%%%%%%%%%%%
\subsection{Clause Scope Refactorings}\label{sub:clause}

The clause scope affects a single clause in a predicate. Usually, this does
not affect any code outside the clause directly. %% An example is when a part
%%of the clause body is replaced by a call to a new predicate that consists of
%%the original goal. Code outside of the clause scope is affected in as much
%%as that the new predicate is introduced and the necessary import declarations
%%are added or removed.  

% \renewcommand{\FancyVerbFormatLine}[1]{%
%         \makebox[3mm][l]{\textbullet}#1}
% \begin{Verbatim}[commandchars=\\\{\},frame=single,fontfamily=\familydefault,framesep=\codeframesep,label=Clause Scope Refactorings,xleftmargin=1cm,xrightmargin=1cm]
% Extract predicate locally
% Invert if-then-else (negate condition and reorder branches)
% Replace cut by if-then-else
% Replace unification by (in)equality test
% Produce output after commit
% \end{Verbatim}
% \renewcommand{\FancyVerbFormatLine}[1]{#1}

\vspace{-5mm}
\subsubsection{Extract predicate locally}

Similarly to the system-scope refactoring with the same name this technique
replaces body subgoals with a call to a new predicate defined by these
subgoals. Unlike for the system-scope here we do not aim to automatically
discover useful candidates for replacement or to replace similar sequences
in the entire system. The user is responsible for selecting the subgoal that
should be extracted.

By restructuring a clause this refactoring technique
can improve its readability. Suitable candidates for this transformation
are clauses with overly large bodies or clauses performing several distinct
subtasks. By cutting the bodies of clauses down to size and isolating subtasks,
it becomes easier for programmers to understand their meaning.

\vspace{-5mm}
\subsubsection{Invert if-then-else}

The idea behind this transformation is that while logically the order of the
``then'' and the ``else'' branches does not matter, it can be important for
code readability. Indeed, an important readability criterion is to have an
intuitive and simple condition. The semantics of the if-then-else construct
in Prolog have been for years a source of controversy \cite{ite:ALPN91}
until it was finally fixed in the ISO standard \cite{ISO13211-1}.  The main
issue is that its semantics differ greatly from those of other programming
languages. Restricting oneself to only conditions that do not bind variables
but only perform tests\footnote{This is similar to the guideline in imperative
languages not to use assignments or other side effects
in conditions.}, makes it easier to understand the meaning of the if-then-else.

To enhance readability it might be worth putting the shorter branch as
``then'' and the longer one as ``else''. Alternatively, the negation of
the condition may be more readable, for example a double negation can be
eliminated. This transformation might also disclose other transformations
that simplify the code.

Hence, we suggest a technique replacing \texttt{(P -> Q ; R)} with
\texttt{($\backslash$+ P -> R ; P, Q)}. Of course, for a built-in \texttt{P}
\vipress generates the appropriate negated built-in instead of \texttt{$\backslash$+ P}.
The call to \texttt{P} in the ``else'' branch is there to keep any bindings
generated in \texttt{P}. If it can be inferred that \texttt{P} cannot generate
any bindings,
%%%% (e.g. because \texttt{P} is a built-in known not to generate any bindings)
then \texttt{P} can be omitted from the ``else'' branch.
\vspace{-5mm}
\subsubsection{Replace cut by if-then-else}

This technique aims at improving program readability by replacing
cuts (!) by if-then-else ({\tt  -> ; }). %%%The cut was introduced in order to prune Prolog's search space
%%%during program execution. 
Despite the controversy on the use of cut inside
the logic programming community,
%%% and recurring attempts to banish it from Prolog~\cite{Debray:Warren,Moss},
it is commonly used in
practical applications both for efficiency and for correctness reasons. We
suggest a transformation that replaces some uses of cut by the more declarative
and potentially more efficient if-then-else.

\begin{example}\label{xmp:fac}

Figure \ref{fig:fac} shows how this refactoring in \vipress transforms 
the program on the left to the program on the right.

%% Our transformation allows us to replace the left-hand side program with the
%% right-hand side program:
%% \begin{Verbatim}[commandchars=\\\{\},frame=single,framesep=5mm,label=If-Then-Else Introduction]
%% fac(0,1) :- !.               fac(N,F) :-
%% fac(N,F) :-                          \underline{(} N = 0, F = 1 \underline{->}
%%         N1 is N - 1,                         true
%%         fac(N1,F1),                  \underline{;}
%%         F is N * F1.                         N1 is N - 1,
%%                                              fac(N1,F1),
%%                                              F is N * F1
%%                                      \underline{)}.
%% \end{Verbatim}

\begin{figure}[h]
\begin{center}
 \subfigure[Before]{
    \includegraphics[width=.38\textwidth]{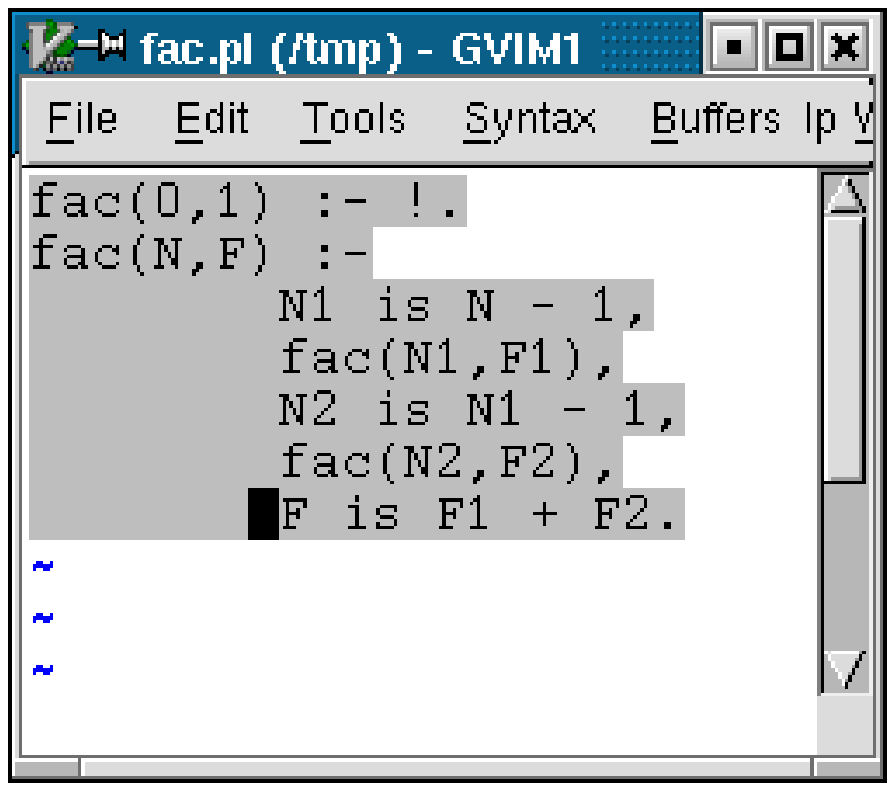}
  }
 \subfigure[After]{
    \includegraphics[width=.38\textwidth]{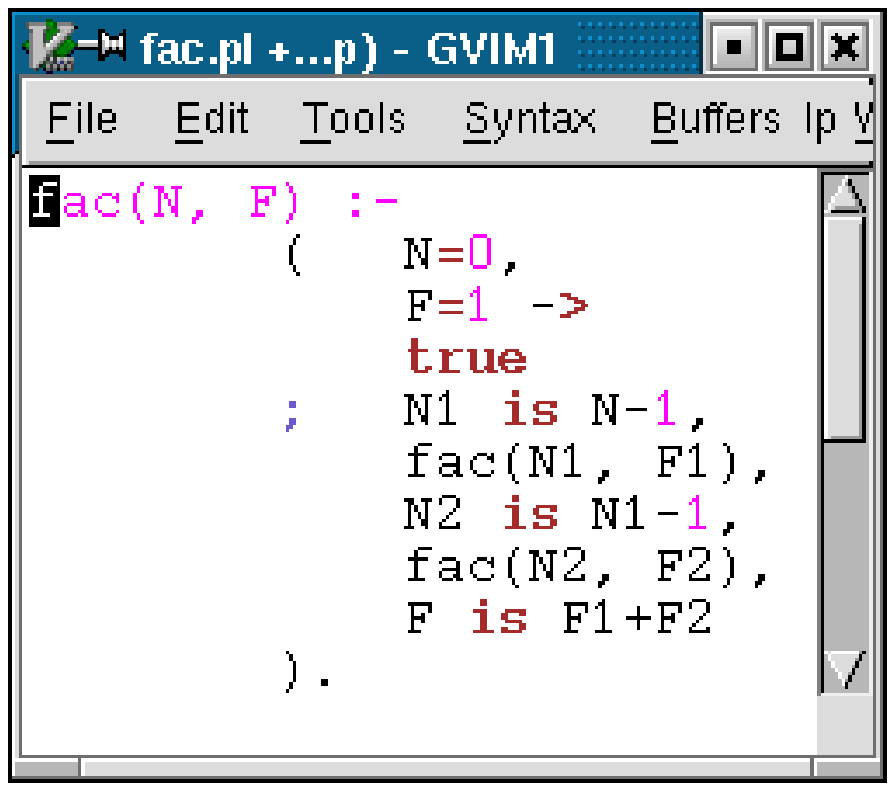}
  }
\end{center}
\caption{\label{fig:fac}Replace cut by if-then-else in \texttt{ViPReSS}.}
\end{figure}

\end{example}

The right-hand side program shows that the refactoring preserves operational
semantics. Moreover, assuming that \texttt{N} is the input and \texttt{F}
the output of \texttt{fac/2}, the refactoring reveals hidden malpractices.
These malpractices are discussed in more detail in the next two refactorings.

\vspace{-5mm}
\subsubsection{Replace unification by (in)equality test}

The previous refactoring
%%%%, {\em replace cut by if-then-else}, 
may expose a hidden
malpractice: full unifications are used instead of equality or other tests.

O'Keefe in \cite{OKeefe} advocates the importance of steadfast code: code that
produces the right answers for all possible modes and inputs. A more moderate
approach is to write code that works for the intended mode only.
%Indeed, Prolog programmers typically write a predicate with a particular mode
%in mind and it is often not easy to see at a glance what that intended mode is.

Unification succeeds in several modes and so does not convey a particular
intended mode. Equality ({\tt ==}, {\tt =:=}) and inequality ({\tt \verb+\==+},
{\tt \verb+=\=+}) checks usually only succeed for one particular mode and
fail or raise an error for other modes. Hence their presence makes it
easier in the code and at runtime to see the intended mode. Moreover, if
only a comparison was intended, then full unification may lead to unwanted
behaviour in unforeseen cases.

The two versions of \texttt{fac/2} in Example \ref{xmp:fac} use unification
to compare \texttt{N} to \texttt{0}. This succeeds if \texttt{N} is variable
by binding it, although this is not the intended mode of the predicate.
By replacing \texttt{N = 0} with \texttt{N == 0} we indicate that \texttt{N}
has to be instantiated to \texttt{0}. This makes it easier for future
maintenance to understand the intended mode of the predicate.  A weaker check
is \texttt{N =:= 0} which allows \texttt{N} to be any expression that evaluates
to \texttt{0}. It may be worthwhile to consider a slightly bigger change of
semantics: \texttt{N =< 0} turns the predicate into a total function. Another
way to avoid an infinite loop for negative input is to add \texttt{N > 0}
to the recursive clause. These checks capture the intended meaning better
than the original unification.

%Note that equality tests are cheaper to execute in some Prolog systems,
%especially if they appear as the only goal in the condition of an if-then-else.
%Nevertheless, the main intent of this refactoring is to bring the operational
%semantics closer to the intended semantics of the programmer. If only a
%comparison is required, then full unification may lead to unwanted behaviour
%in unforeseen cases.
%
%Unfortunately, in most Prolog systems we are aware of, head indexing is lost
%if a unification in the head is replaced by an equality test in the body.
%ECL$^i$PS$^e$
% \cite{ECLIPSE:Manual}
%, however, allows with the \texttt{-?->/1}
%declaration to specify on the clause level that pattern matching and not
%full unification should be used for the head, without loss of indexing.

\vspace{-5mm}
\subsubsection{Produce output after commit}

Another malpractice that may be revealed by the {\em replace cut by
if-then-else} refactoring, is producing output before the commit.
This malpractice is disapproved of by O'Keefe in \cite{OKeefe}, in
line with his advocacy for steadfast predicates.

Now consider what happens with the predicate \texttt{fac/2} in Example
\ref{xmp:fac} if is called as \texttt{?- fac(0,0)}. It does not fail.
On the contrary, it backtracks into the second clause and goes into an
infinite loop. On the other hand, the query \texttt{?- fac(0,F), F=0}
does fail.  Contrary to the intuition which holds for pure Prolog programs,
it is not always valid to further instantiate a query than was intended by
the programmer.

By producing output after the commit, the second clause
can no longer be considered as an alternative for the first query. Hence, the
following version of the first clause has better steadfastness properties:
\texttt{fac(0,F) :- !, F = 1.}
This refactoring may have an impact on the efficiency of the code.
If the output is produced before a particular clause or case is committed to
and this fails, other cases may be tried, which incurs an overhead.
This is illustrated to the extreme with the non-terminating \texttt{fac(0,0)}
query.

\section{The \vipress refactoring browser}\label{sec:vipress}

The refactoring techniques presented above have been implemented in
the refactoring browser \mbox{\vipressnospace}\footnote{Vi(m) P(rolog)
Re(factoring) (by) S(chrijvers) (and) S(erebrenik)}. To facilitate acceptance
of the tool \vipress by the developers community it has been implemented
%%%not as a stand-alone application but 
on the basis of VIM%%%({\tt http://www.vim.org/})
, a popular clone of the well-known VI editor. Techniques like {\em predicate
duplication} provided
 are easy
to implement with the text editing facilities of VIM.

Most of the refactoring tasks have been implemented as SICStus Prolog
\cite{SICStus:Manual} programs inspecting source files and/or call
graphs. Updates to files have been implemented either directly in the
scripting language of VIM or, in the case many files had to be updated
at once, through \texttt{ed} scripts. VIM functions have been written to
initiate the refactorings and to get user input.

% \begin{figure}[h]
% \begin{center}
% \includegraphics[width=\textwidth]{extract_predicate}
% \end{center}
%     \caption{Screenshot of ViPReSS in action: {\em extract predicate locally}\label{fig:screenshot}}
% \end{figure}

% Figure \ref{fig:screenshot} shows a screenshot of {\em extract predicate
% locally} in VIM. The user selects the subgoals that are to be extracted into
% a predicate and then invokes the refactoring by hitting the appropriate
% key. Then the user enters the desired predicate name. Finally, the file
% is filtered through a Prolog program that generates the new predicate and
% replaces the original goals by a call to it.

\vipress has been successfully applied to a large (more than 53,000 lines)
legacy system used at the Computer Science department of the Katholieke
Universiteit Leuven to manage the educational activities. The system, called
\btwnospace, %%%%% (Flemish for value-added tax)
has been developed and extended since the
early eighties by more than ten different programmers, many of whom are no
longer employed by the department. The implementation has been done in the
MasterProLog~\cite{MasterProLog} system that, to the best of our knowledge,
is no longer supported.

By using the refactoring techniques we succeeded in obtaining a better
understanding of this real-world system, in improving its structure and
maintainability, and in preparing it for further intended changes such
as porting it to a state-of-the-art Prolog system and adapting it to
new educational tasks the department is facing as a part of the unified
Bachelor-Master system in Europe.

We started by removing some parts of the system that have been identified
by the expert as obsolete, including out-of-fashion user interfaces and
outdated versions of program files. The bulk of dead code was eliminated in
this way, reducing the system size to a mere 20,000 lines. 
% Note that dead
% code elimination after simply removing the outdated toplevel predicates
% would have had the same effect.

Next, we applied most of the system-scope refactorings described above. Even
after removal of dead code by the experts \vipress identified and eliminated
299 dead predicates. This reduced the size by another 1,500 lines.  Moreover
\vipress discovered 79 pairwise identical predicates. In most of the cases,
identical predicates were moved to new modules used by the original ones. The
previous steps allowed us to improve the overall structure of the program
by reducing the number of files from 294 to 116 files with a total of 18,000
lines. Very little time was spent to bring the system into this state. The
experts were sufficiently familiar with the system to immediately identify
obsolete parts. The system-scope refactorings took only a few minutes each.

The second step of refactoring consisted of a thorough code inspection aimed
at local improvement. Many malpractices have been identified: excessive
use of cut combined with producing the output before commit being the
most notorious one. Additional ``bad smells'' discovered include bad
predicate names such as {\tt q}, unused arguments and unifications
instead of identity checks or numerical equalities. Some of these were
located by \vipress, others were recognised by the users, while \vipress
performed the corresponding transformations. This step is more
demanding of the user. She has to consider all potential candidates for
refactoring separately and decide on what transformations apply. Hence,
the lion's share of the refactoring time is spent on these local changes.

In summary, from the case study we learned that automatic support for
refactoring techniques is essential and that \vipress is well-suited
for this task. As the result of applying refactoring to \btw we obtained
better-structured lumber-free code. Now it is not only more readable and
understandable but it also simplifies implementing the intended changes. From
our experience with refactoring this large legacy system and the relative
time investments of the global and the local refactorings, we recommend to
start out with the global ones and then selectively apply local refactorings
as the need occurs.

A version of \vipress to refactor SICStus programs can be downloaded from
\texttt{http://www.cs.kuleuven.ac.be/\~{}toms/vipress}. The current version,
0.2.1, consists of 1,559 lines of code and can also refactor ISO Prolog
programs. Dependencies on the system specific builtins and the module system
have been separated as much as possible from the refactoring logic. This
should make it fairly easy to refactor other Prolog variants as well.

%%%%%%%%%%%%%%%%%%%%%%%%%%%%%%%%%%%%%%%%%%%%%%%%%%%%%%%%%%%%%%%%%%%%%%%%%%%%%%%%
\section{Conclusions and Future Work}\label{sec:conclusions}

In this paper we have shown that the ideas of refactoring
are applicable and important for logic programming. Refactoring
helps bridging the gap between prototypes and real-world applications. Indeed,
extending a prototype to provide additional functionality often leads to
cumbersome code. Refactoring allows software developers both to clean up
code after changes and to prepare code for future changes.
 
We have presented a catalogue of refactorings, %%%% at different scopes of a
%%%program, 
containing both previously known refactorings for object-oriented
languages now adapted for Prolog and entirely new Prolog-specific
refactorings. Although the presented refactorings do require human input
as it is in the general spirit of refactoring, a
large part of the work can be automated. Our refactoring browser \vipress
integrates the automatable parts of the presented refactorings in the VIM
editor.
%  and a version to refactor SICStus programs can be downloaded from
% \texttt{http://www.cs.kuleuven.ac.be/\~{}toms/vipress}.

Logic programming languages and refactoring have already been put together
at different levels. Tarau~\cite{Tarau:Refactoring} has refactored the
Prolog language itself. However, this approach differs significantly
from the traditional notion of refactoring %%%as introduced by Fowler
~\cite{Fowler:et:al}. We follow the latter definition.  Recent relevant work
is~\cite{Tourwe:Mens} in the context of object oriented languages: a meta-logic
very similar to Prolog is used to detect for instance obsolete parameters.

None of these papers, however, considers applying refactoring techniques
to logic programs. 
% In our previous work~\cite{Serebrenik:Demoen} we have
% emphasised the importance of refactoring for logic programming and discussed
% the applicability of the refactoring techniques developed for object-oriented
% languages to Prolog and CLP-programs. 
Seipel {\em et
al.}~\cite{Seipel:Hopfner:Heumesser} include refactoring among the analysis
and visualisation techniques that can be easily implemented by means of {\sc
FnQuery}, a Prolog-inspired query language for XML. However, the discussion
stays at the level of an example and no detailed study has been conducted.

In the logic programming community questions related to refactoring have been
intensively studied in context of program transformation and specialisation
\cite{Deville,Etalle:Gabbrielli:Meo,Leuschel:Sorensen,Pettorossi:Proietti}.
There are two important differences with this line of work. Firstly,
refactoring does not aim at optimising performance but at improving
readability, maintainability and extensibility. In the past these
features where often sacrified to achieve efficiency. Secondly, user input
is essential in the refactoring process while traditionally only automatic
approaches were considered. Moreover, usually program transformations
are part of a compiler and hence, they are ``invisible'' to the program
developer. However, some of the transformations developed for program
optimisation, e.g. {\em dead code elimination}, can be considered as
refactorings and should be implemented in refactoring browsers.

To further increase the level of automation of particular refactorings
additional information such as types and modes can be used.  To obtain
this information the refactoring system could be extended with type and
mode analyses.
On the other hand, it seems worthwhile to consider the proposed refactorings
in the context of languages with type and mode declarations like Mercury
\cite{Mercury}, especially as these languages claim to
be of greater relevance for programming in the large than traditional Prolog.
Moreover, dealing with higher order features is essential for refactoring in a real
world context. The above mentioned languages with explicit declarations for
such constructs would facilitate the implementation of an industrial
strength refactoring environment.
\vspace{-2mm}

%%%%%%%%%%%%%%%%%%%%%%%%%%%%%%%%%%%%%%%%%%%%%%%%%%%%%%%%%%%%%%%%%%%%%%%%%%%%%%%%

\bibliography{paper}

\begin{thebibliography}{10}

\bibitem{ite:ALPN91}
The {\tt ->} operator.
\newblock {\em Association for Logic Programming Newsletter}, 4(2):10--12,
  1991.

\bibitem{ISO13211-1}
{\em Information technology---Programming languages---{P}rolog---Part 1:
  General core}.
\newblock {ISO/IEC}, 1995.
\newblock {{ISO/IEC}} 13211-1:1995.

\bibitem{Deville}
Y.~Deville.
\newblock {\em Logic Programming: Systematic program development}.
\newblock Addison-Wesley, 1990.

\bibitem{Etalle:Gabbrielli:Meo}
S.~Etalle, M.~Gabbrielli, and M.~C. Meo.
\newblock Transformations of {CCP} programs.
\newblock {\em {ACM} {T}ransactions on {P}rogramming {L}anguages and
  {S}ystems}, 23(3):304--395, May 2001.

\bibitem{Fowler:catalogue}
M.~Fowler.
\newblock Refactorings in alphabetical order.
\newblock Available at \verb+http://www.refactoring.com/catalog/+, 2003.

\bibitem{Fowler:et:al}
M.~Fowler, K.~Beck, J.~Brant, W.~Opdyke, and D.~Roberts.
\newblock {\em Refactoring: improving the design of existing code}.
\newblock Object Technology Series. Addison-Wesley, 1999.

\bibitem{SICStus:Manual}
{Intelligent Systems Laboratory}.
\newblock {\em {SICStus Prolog User's Manual}}.
\newblock {PO Box 1263, SE-164 29 Kista, Sweden}, October 2003.

\bibitem{MasterProLog}
{IT M}asters.
\newblock Master{P}ro{L}og {P}rogramming {E}nvironment.
\newblock \verb+www.itmasters.com+, 2000.

\bibitem{Jacobs:Blockeel}
N.~Jacobs and H.~Blockeel.
\newblock The learning shell : Automated macro construction.
\newblock In {\em User Modeling 2001}, volume 2109 of {\em LNAI}, pages 34--43.
  Springer Verlag, 2001.

\bibitem{Leuschel:Sorensen}
M.~Leuschel and M.~H. S{\o}rensen.
\newblock Redundant argument filtering of logic programs.
\newblock In J.~Gallagher, editor, {\em Proceedings of the $6^{th}$
  International Workshop on Logic Program Synthesis and Transformation}, volume
  1207 of {\em LNCS}, pages 83--103. Springer Verlag, 1996.

\bibitem{Li:Reinke:Thompson}
H.~Li, C.~Reinke, and S.~Thompson.
\newblock Tool support for refactoring functional programs.
\newblock In J.~Jeuring, editor, {\em {H}askell Workshop 2003}. Association for
  {C}omputing {M}achinery, 2003.

\bibitem{OKeefe}
R.~A. O'Keefe.
\newblock {\em The {C}raft of {P}rolog}.
\newblock {MIT} {P}ress, Cambridge, {MA}, {USA}, 1994.

\bibitem{Opdyke:PhD}
W.~F. Opdyke.
\newblock {\em Refactoring object-oriented frameworks}.
\newblock PhD thesis, University of Illinois at Urbana-Champaign, 1992.

\bibitem{Pettorossi:Proietti}
A.~Pettorossi and M.~Proietti.
\newblock Transformation of logic programs: Foundations and techniques.
\newblock {\em Journal of Logic Programming}, 19/20:261--320, May/July 1994.

\bibitem{Roberts:Brant:Johnson}
D.~Roberts, J.~Brant, and R.~Johnson.
\newblock A refactoring tool for {S}malltalk.
\newblock {\em Theory and {P}ractice of {O}bject{S}ystems ({TAPOS})},
  3(4):253--263, 1997.

\bibitem{Schrijvers:Serebrenik:Demoen}
T.~Schrijvers, A.~Serebrenik, and B.~Demoen.
\newblock Refactoring {Prolog} programs.
\newblock Technical Report CW373, Department of Computerscience, K.U.Leuven,
  December 2003.

\bibitem{Seipel:Hopfner:Heumesser}
D.~Seipel, M.~Hopfner, and B.~Heumesser.
\newblock Analysing and visualizing {P}rolog programs based on {XML}
  representations.
\newblock In F.~Mesnard and A.~Serebrenik, editors, {\em Proceedings of the
  13th International Workshop on Logic Programming Environments}, pages 31--45,
  2003.
\newblock Published as technical report {CW}371 of {K}atholieke {U}niversiteit
  {L}euven.

\bibitem{Mercury}
Z.~Somogyi, F.~Henderson, and T.~Conway.
\newblock Mercury: an efficient purely declarative logic programming language.
\newblock In {\em Australian Computer Science Conference}.

\bibitem{Tarau:Refactoring}
P.~Tarau.
\newblock Fluents: A refactoring of {P}rolog for uniform reflection an
  interoperation with external objects.
\newblock In {\em Computational Logic, First International Conference, London,
  UK, July 2000, Proceedings}, volume 1861 of {\em LNAI}, pages 1225--1239.
  Springer Verlag, 2000.

\bibitem{Tourwe:Mens}
T.~Tourw{\'e} and T.~Mens.
\newblock Identifying refactoring opportunities using logic meta programming.
\newblock In {\em 7th European Conference on Software Maintenance and
  Reengineering, Proceedings}, pages 91--100. IEEE Computer Society, 2003.

\end{thebibliography}

\bibliographystyle{abbrv}

%%%%%%%%%%%%%%%%%%%%%%%%%%%%%%%%%%%%%%%%%%%%%%%%%%%%%%%%%%%%%%%%%%%%%%%%%%%%%%%%
\end{document}